# Internal dynamics and fission of pure-quartic soliton molecules

Zhixiang Deng,[1] Rui Ma,[1] Chunxiang Zhang,[2] Boris Malomed,[3] Dianyuan Fan,[1] Jingsong He,[4] and Jun Liu[1,*]

[1]*International Collaborative Laboratory of 2D Materials for Optoelectronics Science & Technology, Institute of Microscale Optoelectronics, Shenzhen University, Shenzhen 518060, China*

[2]*Shenzhen Key Laboratory of Ultraintense Laser and Advanced Material Technology, Center for Advanced Material Diagnostic Technology, College of Engineering Physics, Shenzhen Technology University, Shenzhen 518118, China*

[3]*Department of Interdisciplinary Studies, Faculty of Engineering, Tel Aviv University, Tel Aviv 69978, Israel*

[4]*Institute for Advanced Study, Shenzhen University, Shenzhen 518060, China*

We address the weak interaction of a pair of well-separated pure-quartic solitons (PQSs), which are solutions to a generalized nonlinear Schrödinger equation (NLSE) with the quartic-only dispersion. An asymptotic technique is applied to derive equations for the slow evolution of the temporal separation and phase difference of the PQSs interacting through the overlapping of their exponentially decaying oscillating tails. Based on this approach, various stationary states of bound PQS ("soliton molecules") with distinct phase differences are predicted. Their stability is addressed via the numerical calculation of the eigenvalue spectrum of small perturbations, showing instability of the bound states. A systematic numerical analysis demonstrates that the parameter space of the PQS bound states is organized as a self-similar fractal structure, composed of regions populated by robustly oscillating or splitting two-soliton states. The analytical method and results reported here can be extended for bound states of two or several weakly interacting modes in other conservative and dissipative systems.

## I. INTRODUCTION

It is commonly known that optical solitons, which are represented by localized solutions to nonlinear Schrödinger equations (NLSEs), are maintained by the balance between the diffraction/dispersion and nonlinearity. Conventional solitons remain stable under the action of perturbations [1, 2], which are highly beneficial for applications (first of all, to optical-data transmission and processing [3]), also providing a fundamental paradigm for studying localized-wave dynamics in optics and other areas of physics. Besides isolated solitons, the Schrödinger-like nonlinear systems can support multi-soliton complexes [4-6]. Well-separated solitons can form bound states (also called "soliton molecules" [7]) through their weak interactions mediated by exponentially decaying oscillatory tails [8]. The dynamics of soliton bound states have been studied in a great variety of nonlinear systems [9-24]. In particular, in nonlinear optical setups governed by the complex Ginzburg-Landau equations (CGLEs) [12-18] and Lugiato-Lefever equations (LLEs) [19-22, 25], the dynamics of stable bound complexes of dissipative solitons were investigated theoretically and experimentally. Note, however, that the second-order NLSEs (including the classical integrable equation), which produce traditional soliton solutions, do not give rise to oscillatory tails of soliton solutions; hence this basic equation does not support the existence of stationary bound states of well-separated solitons [1-3].

Pure-quartic solitons (PQSs), proposed and experimentally realized in recent works [26-29], are represented by localized solutions of a non-integrable NLSE with the self-focusing nonlinearity balanced by the negative quartic dispersion, in the absence of the usual second-order dispersion term. Basic characteristics of PQSs include two aspects: One is the energy-width scaling relation [26, 30], which indicates that the energy of PQSs can be scaled to several orders of magnitude higher than for the standard NLSE solitons with a similar pulse duration. This advantage provides strong stability of the solitons against random jitter and suggests new possibilities to design high-energy soliton fiber lasers [30]. Another important feature of the PQSs is the oscillatory shape of their exponentially decaying tails [26, 31]. The latter feature, occurring in the conservative system, unlike CGLE, naturally facilitates the formation of bound state of doubly solitons, without their coalescence. Recently, interactions and collisions of PQSs were explored numerically, referring to an effective interaction potential of the two-soliton state [31-34]. However, the full investigations of weak interactions and oscillatory motions between their internal constituents, including their mutual phase locking, have not been performed yet.

In this paper, the investigation of weak interactions between well-separated PQSs is organized as follows. In Sec. II, we outline the typical behavior of the PQSs tails, and derive the interaction equations for the slow evolution of their temporal positions and relative phase. In Sec. III, the equilibrating dynamics for the two-soliton bound states are predicted by the interaction model, and their stability is explored by calculating the linear eigenvalue spectra for small perturbations. In Sec. IV, we turn to the oscillating dynamics for the bound-PQS states around the equilibrium position, illustrating a self-similar structure of the weak interaction of well-separated PQSs. The temporal separations of the double-PQS pairs display a fractal behavior as the initial spacing is varied. Conclusions are summarized in Sec. V. The detailed derivation of the PQS interaction equations is illustrated in Appendix A while the corresponding analysis based on the conserved quantity is given in Appendix B.

*[liu-jun-1987@live.cn](mailto:liu-jun-1987@live.cn)

## II. THEORY OF PQSs INTERACTION

The generalized NLSE for complex amplitude $u$ of the optical field, with coefficients $\beta_4$ and $\gamma$ of the pure quartic dispersion and cubic self-focusing nonlinearity, is written as [26-28]:

$$i\frac{\partial u}{\partial z} - \frac{|\beta_4|}{24}\frac{\partial^4 u}{\partial t^4} + \gamma |u|^2 u = 0, \qquad (1)$$

where $z$ and $t$ are the propagation distance and the retarded time, respectively. By separating the real and imaginary parts of the complex amplitude, $u = u_1 + iu_2$, Eq. (1) can be cast in the vectorial form for $\mathbf{U} = (u_1, u_2)^T$,

$$\partial_z \mathbf{U} = \mathbf{F}(\mathbf{U}), \qquad (2)$$

$$\mathbf{F}(\mathbf{U}) = \frac{|\beta_4|}{24}\mathbf{J}\cdot\partial_t^4 \mathbf{U} - \gamma(u_1^2 + u_2^2)\mathbf{J}\cdot\mathbf{U}, \qquad (3)$$

with $\mathbf{F}(\mathbf{0}) = \mathbf{0}$, and the symplectic matrix $\mathbf{J} = \begin{pmatrix} 0 & 1 \\ -1 & 0 \end{pmatrix}$.

### A. PQSs tails

Stationary solutions to Eq. (1) are sought for as $A(t)e^{i\mu z}$ with propagation constant $\mu > 0$, which satisfies the ordinary differential equation (ODE) $-\mu A - \frac{|\beta_4|}{24}\frac{d^4 A}{dt^4} + \gamma A^3 = 0$. While the analytical PQS solutions of this equation are not available, they can be solved numerically, using the Newton's conjugate-gradient (NCG) method [35].

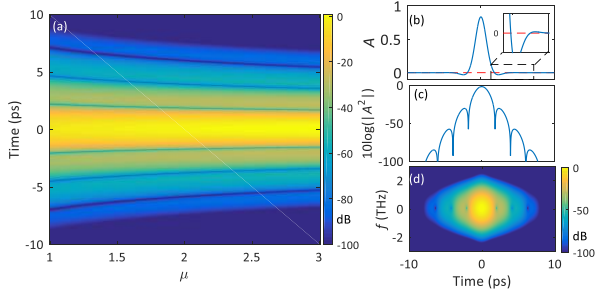

FIG. 1. Numerically found PQS solutions, i.e., stationary solutions of Eq. (1). (a) The intensity profiles of PQSs on a logarithmic scale as a function of the propagation constant $\mu$. (b) The amplitude profile of PQS for $\mu=1.76$ mm$^{-1}$ and (c) the corresponding intensity profile, plotted on a logarithmic scale. (d) The spectrogram of PQS at $\mu=1.76$ mm$^{-1}$. Physical parameters in Eq. (1) are the same as in Ref. [28]: $\beta_4=-2.2$ ps$^4$mm$^{-1}$ and $\gamma=4.072$ W$^{-1}$mm$^{-1}$.

Figure 1(a) presents the numerically calculated intensity evolution of PQSs as a function of the propagation constant $\mu$, with profiles of the absolute value of the amplitude and intensity plotted for $\mu=1.76$ mm$^{-1}$ in Figs. 1(b) and 1(c) on the linear and logarithmic scales, respectively. Recent studies [26-28] have found that the PQSs do not feature a smooth (hyperbolic-secant) shape of the traditional NLSE solitons. Instead, they exhibit exponentially decaying oscillatory tails on both sides of the PQS's core. The decay rate and oscillating period of the tails are determined by the linearization of ODE, $-\mu A - \frac{|\beta_4|}{24}\frac{d^4 A}{dt^4} = 0$. Assuming that the solution for the PQS tails is $A = \sum_{j=1}^{4} a_j e^{\lambda_j t}$ with constant $a_j$, one obtains the eigenvalues $\lambda = \pm G(1 \pm i)$ with $G = (6\mu/|\beta_4|)^{1/4}$ and two independent sign sets $\pm$. Obviously, the eigenvalues with positive and negative real parts pertain to the leading and trailing tails of the PQS, respectively. Thus, far from the PQS's core, its tails take the asymptotic form

$$A_0(t) \approx ae^{-G(1\pm i)|t|}, \qquad (4)$$

where $a$ is a complex constant determined by matching the tails to the core, which should be found numerically. The approximate expression is then written as $A_0(t) \approx K \cdot \sqrt{\frac{\mu}{\gamma}} e^{-G|t|} \cos(G|t| + \theta)$, which the constant $K = 4.48209$, $\theta = -1.16201$. It is seen from Fig. 1(a) that, for a large range of the values $\mu$, the tails indeed exhibit the exponential decay with visible oscillations spreading sufficiently far from the soliton's core. This is a fundamental feature which underlies the weakly interaction between separated PQSs [8].

Furthermore, the asymptotic form in the PQS tails is also illustrated in Fig. 1(d) by the spectrogram, defined as $P(t, f) = |\int_{-\infty}^{\infty} A(t') \exp[-2(t'-t)^2 - i2\pi f \cdot t']dt'|^2$. It features a few equidistant holes in the temporal direction at frequency $f = 0$, which confirms the oscillatory shape of the PQS tails.

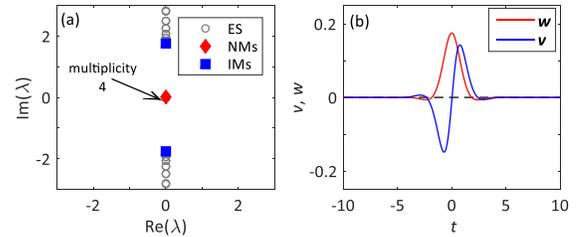

FIG. 2. The spectra of PQSs (a) and its neutral (Goldstone) modes $w$, $v$ (b). In (a), acronyms ES, NMs, and IMs mean, respectively, the essential spectrum, neutral mode eigenvalues, and internal mode eigenvalues.

### B. The equation of weak interaction for well-separated PQSs

To describe the interaction of far separated PQSs, one needs to derive the equations governing the slow evolution of the temporal positions and phases of individual PQSs. First, we perform the linear stability analysis of an isolated stationary PQSs, $\mathbf{U} = \mathbf{u}_0$, which is determined by the spectrum of operator $\mathbf{L}$ derived from the linearization of Eq. (2) on this solution. The numerically found spectrum is presented in Fig. 2(a), which establishes the spectral stability of the PQS, and is in agreement with Ref. [5, 28]. The spectrum includes the essential (continuous) spectrum, internal-mode eigenvalues, and zero ones, cf. Ref. [28]. We focus here on the eigenvalues at the origin, which have an algebraic multiplicity of 4. Two of them correspond to the phase-shift and translational symmetries of Eq. (1), i.e., $u(t) \to u(t)e^{i\phi_0}$ and $u(t) \to u(t-t_0)$ with arbitrary real constants $t_0, \phi_0$. The respective neutral (Goldstone) eigenmodes of the linearization operator $\mathcal{L}$ are $\mathbf{w} = i\mathbf{u}_0$ and $\mathbf{v} = \partial_t \mathbf{u}_0$, respectively. They are presented in Fig. 2(b). The two generalized discrete eigenfunctions $\mathbf{w}_g, \mathbf{v}_g$, corresponding to the other two eigenvalues at the origin, can be determined by numerically calculating the linear equations $\mathcal{L}\mathbf{w}_g = \mathbf{w}, \mathcal{L}\mathbf{v}_g = \mathbf{v}$. Along with the operator $\mathcal{L}$ for the stable PQS, we will use its adjoint operator $\mathcal{L}^\dagger$ such that $\mathcal{L}^\dagger = \sigma_3 \mathcal{L} \sigma_3^{-1}$, where $\sigma_3 = \text{diag}(1,-1)$. It also can be defined by the relation $\langle \mathbf{v}, \mathcal{L}\mathbf{w}\rangle = \langle \mathcal{L}^\dagger \mathbf{v}, \mathbf{w}\rangle$, where the inner product is $\langle \mathbf{v}, \mathbf{w}\rangle = 2\int_{-\infty}^{\infty} \text{Re}[v^*(x)w(x)]dx$. Therefore, the adjoint neutral eigenmodes corresponding to the phase-shift and translational symmetries of Eq. (1) are $\mathbf{w}^\dagger = \sigma_3 \mathbf{w}$ and $\mathbf{v}^\dagger = \sigma_3 \mathbf{v}$, respectively. Thus, at sufficiently large distances from the PQS core, the asymptotic form of the adjoint neutral modes is:

$$v_0(t) \approx b\, \text{sgn}(t)\cdot e^{-G(1\pm i)|t|}, \quad w_0(t) \approx c e^{-G(1\mp i)|t|}, t \to \pm\infty, \quad (5)$$

with complex constants $b$ and $c$ that, like $a$ in Eq. (4), should found numerically from the matching to the PQS core.

Next, we search for double-peak solutions of Eq. (2) in the form of a stationary bound state ("soliton molecules") of two well-separated PQSs, cf. Ref. [8, 10, 14, 22],

$$\mathbf{U} = \mathbf{u}_\Sigma + \mathbf{u}^{(1)} + \mathbf{u}^{(2)}, \quad \mathbf{u}_\Sigma = \mathbf{u}_1 + \mathbf{u}_2, \quad (6)$$

where $\mathbf{u}_{1,2} = \mathbf{u}_0(t-\tau_{1,2})e^{i\phi_{1,2}}$ and the corrections $\mathbf{u}^{(m)} = O(\varepsilon^m), m = 1, 2$. Here the parameter $\varepsilon$ characterizes the weakness of the overlapping between tails of the interacting PQSs. The temporal positions $\tau_{1,2}$ and the phases $\phi_{1,2}$ of the PQSs are slowly varying functions of the propagation distance, with a slowness estimates $\partial_z \tau_{1,2}, \partial_z \phi_{1,2} = O(\varepsilon)$ and $\partial_{zz}\tau_{1,2}, \partial_{zz}\phi_{1,2} = O(\varepsilon^2)$. By substituting expression (6) in Eq. (2), collecting the terms in $\varepsilon$ and $\varepsilon^2$, and applying the respective solvability conditions [36], we obtain

$$N_v \ddot\tau_{1,2} = -\langle \mathbf{v}^\dagger_{1,2}\cdot\mathbf{P}\rangle, \quad N_w \ddot\phi_{1,2} - N_{1vw}(\dot\tau_{1,2})^2 = \langle \mathbf{w}^?_{1,2}\cdot\mathbf{P}\rangle, \quad (7)$$

$$\mathbf{P} = \partial_z \mathbf{u}_\Sigma - \mathbf{F}(\mathbf{u}_\Sigma), \quad (8)$$

where $\mathbf{v}^\dagger_{1,2} = \mathbf{v}^\dagger(t-\tau_{1,2})e^{i\phi_{1,2}}$, and $\mathbf{w}^\dagger_{1,2} = \mathbf{w}^\dagger(t-\tau_{1,2})e^{i\phi_{1,2}}$. The dots stand for $\partial_z$. The coefficients $N_v, N_w$, and $N_{1vw}$ in Eq. (7) involve the inner products of the generalized eigenfunctions $\mathbf{w}_g, \mathbf{v}_g$ (or its derivative $\partial_t \mathbf{v}_g$) and its adjoint neutral modes $\mathbf{w}^\dagger, \mathbf{v}^\dagger$ that should be calculated, in general, numerically. The derivative process of Eq. (7) describing the weak interactions between far separated PQSs is presented in Appendix A.

In the second-order approximation, the resulting system of slow-evolution equations for the relative parameters of the two-soliton set, $\Delta\tau = \tau_2 - \tau_1$ and $\Delta\phi = \phi_2 - \phi_1$, can be derived analytically, see details in Appendix A:

$$\Delta\ddot\tau \approx 2M_0 e^{-G\cdot\Delta\tau}\cos(\Delta\phi)\sin(G\cdot\Delta\tau+\Theta), \quad (9)$$

$$\Delta\ddot\phi - Q\cdot\dot\tau_{21} \approx -2N_0 e^{-G\cdot\Delta\tau}\sin(\Delta\phi)\cos(G\cdot\Delta\tau+\Xi), \quad (10)$$

where $\dot\tau_{21} = (\dot\tau_2)^2 - (\dot\tau_1)^2$, and the constants are defined as follows: $M_0 N_v e^{i\Theta}/b^* = N_0 N_w e^{i\Xi}/c^* = \sqrt{2}/6\times |\beta_4|aG^3 e^{-i3\pi/4}$, $Q = N_{1vw}/N_v$.

## III. THE EXISTENCE AND STABILITY OF STATIONARY BOUND STATES

Steady-state solutions (fixed points or equilibria) of Eqs. (9) and (10) predict an infinite set of the temporal separation at which the stationary two-soliton bound states ($\dot\tau_1 = \dot\tau_2 = 0$) exist. These are:

$$\begin{aligned} S_0: &\quad \Delta\tau_{eq} = (n\pi - \Theta)/G, \Delta\phi_{eq} = 0; \\ S_{\pm\pi}: &\quad \Delta\tau_{eq} = (n\pi - \Theta)/G, \Delta\phi_{eq} = \pm\pi; \quad (11) \\ S_{\pm\frac{\pi}{2}}: &\quad \Delta\tau_{eq} = [(n+\tfrac{1}{2})\pi - \Xi]/G, \Delta\phi_{eq} = \pm\tfrac{\pi}{2}, \end{aligned}$$

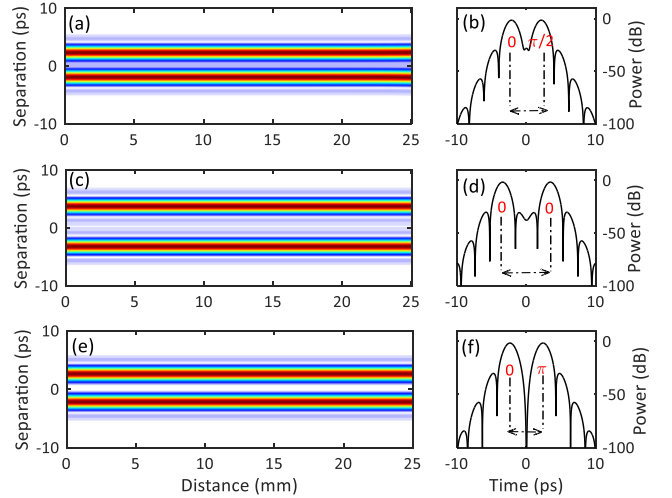

FIG. 3. Typical examples of the bound states of two interacting PQSs ("soliton molecules"), showing the stationary evolution of their double-humped structures. The local-amplitude evolution of the interacting PQSs and the corresponding profiles of the bound states are displayed in the left and right columns, respectively. The phase structures of the bound PQSs are mutually orthogonal in (a, b), in-phase in (c, d), and out-of-phase in (e, f), corresponding, respectively, to the bound states of types $S_{\pm\frac{\pi}{2}}$, $S_0$, and $S_{\pm\pi}$, as defined in Eq. (11).

at $n > n_0$, where $n_0$ corresponds to the bound state with the minimal separation. In reality, the integer $n$ in Eq. (11) is also bound from above, as the soliton pairs with very weak binding, corresponding to a very large separation, may be easily destroyed by external perturbations, such as the quantum noise [37]. Typical evolution pictures for the three classes of the bound states produced by Eq. (11) are presented in Fig. 3. The boundary conditions used in our calculations are periodic. They play a minor role as we deal with the localized PQS solutions decaying to zero as $|t|\to\infty$. Furthermore, we have verified that the dynamical invariants, *viz*, the mass, momentum, and Hamiltonian remain nearly constant in the course of the evolution, with the weak non-conservation (the absolute errors over the evolution processes may be up to $10^{-4}$) induced by the accumulation of radiation effects. The left and right columns of the figure display, severally, the spatial evolution of the pattern of $|u(t,z)|$ in respective doubly-humped solutions, illustrating the motionless behavior along the propagation paths, and the corresponding intensity profiles on a logarithmic scale, produced by means of the NCG method. Specifically, Figs. 3(a) and 3(b) represent the bound state of the type of phase-orthogonal PQSs ($S_{\pm\pi/2}$), while the bound states of in-phase ($S_0$) and out-of-phase ($S_{\pm\pi}$) types are represented by the second and third rows in Fig. 3, respectively. Note that the temporal separation between the soliton's centers is different in the bound states of the different types in Fig. 3. In particular, the maximum and minimum separations, $\Delta\tau$ = 4.85ps and 2.45ps correspond to the in- and out-of-phase bound states, respectively.

An example of the effective interaction force determined by the Eqs. (9) and (10), which corresponds to the phase shift $\Delta\varphi=\pi$ between the solitons, is plotted in Fig. 4(a). Zero points of this profile determine the separation between the solitons in the bound states above (see also the inset in Fig. 4(a)).

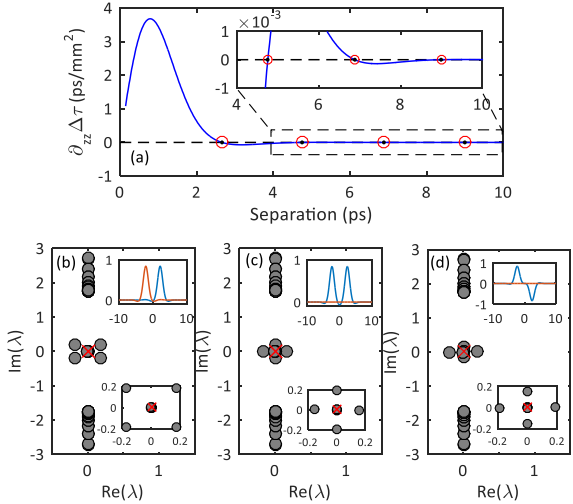

FIG. 4. (a) The right-hand side of Eq. (9) for $\Delta\varphi=\pi$ as a function of the temporal separation $\Delta\tau$ between the interacting PQSs. Red dots indicate fixed points, corresponding to the separation between the solitons in the unstable or stable bound states. (b-d): Stability-eigenvalue spectra corresponding to the three cases displayed in Fig. 3, with upper insets representing the real and imaginary components in the profiles of the stationary bound states and lower insets showing the close-up of the eigenvalues near the origin.

A critically important idea is stability of these bound-PQS structures. To this end, the linear stability analysis is performed based on the numerical calculation of the eigenvalues $\lambda$ of the linearized operator $\mathcal{L}$ associated with Eq. (2) for each particular bound state found in the previous section. The instability is accounted for by the eigenvalues with a positive real part. The corresponding eigenvalue spectra are plotted in Figs. 4(b-d), showing a finite set of eigenvalues near the origin. The eigenvalues at the origin (indicated by a cross) correspond to the phase shift and translational symmetries of stationary bound states. Near the origins, the two pairs of eigenvalues result from the interaction induced by the overlapping between the tails of PQSs. As shown in the lower insets in Fig. 4(b-d), these eigenvalues for the $S_{\pm\pi/2}$ type of bound states appear in the form of quartet $\lambda\approx\pm0.2\pm0.2i$ while they are either real or pure imaginary for the $S_0$ or $S_{\pm\pi}$ type of bound states. Obviously, all three types of the bound states are linearly unstable, since each case has an eigenvalue with positive real part. Nevertheless, it is demonstrated below that, in most cases, the instability for the $S_0$ or $S_{\pm\pi}$ type of bound states does not destroy the "PQS molecules", but rather transforms them into quasi-periodic configurations, which may seem as quite robust oscillatory states (although in some other conditions the "molecules" eventually suffer fission into separating solitons).

## IV. SELF-SIMILAR INTERACTION OF OSCILLATING PQS PAIRS

Next we address the weak interaction of both PQSs beyond the regime of formation of stationary pairs "PQS molecules". Indeed, the instability of the stationary bound states transforms them into a dynamical form. Actually, the simulated dynamics of the bound-PQSs states exhibits quasi-harmonic oscillations (like a recurrence of bounces) around the equilibrium point. Slow oscillations of the "PQS molecules" may be also caused by a deviation of the input from the numerically exact bound state.

### A. Quasi-periodic dynamics of the bound states

First, we concentrate on the formation of bound states of PQSs in the case when the initial separation between them is different from the values corresponding to the numerically exact "QPS molecule". The simulated evolution of the in-phase and out-of-phase QPSs pairs, corresponding, in physical units, to the propagation distance 25mm or 20mm is shown in Fig. 5, which demonstrates the oscillatory evolution of the bound states of the interacting PQSs exhibiting periodic bounce (see also in the insets in Fig. 5). At this point, the conserved quantities in terms of the mass and momentum maintain almost the same during the evolution processes. It should be noted that the Hamiltonians (defined as $H=\int_{-\infty}^{\infty}[-\frac{|\beta_4|}{24}|\partial_t^2 u|^2-\gamma|u|^4]\mathrm{d}t$) exhibit some different oscillation behaviors from that of the stationary bound states since the kinetic contribution is not included. The in-phase bound state exhibits the periodic dynamics in which the interacting PQSs first bounce from each other and then return to the original positions. The evolution is reversed for the out-of-phase bound state, in which the observed interaction periodically switches from attraction to repulsion (bouncing). In other words, the motion of the bound PQSs is mutually synchronous in both, but in opposite directions. Their separation between the solitons oscillates around distinct stable centers (4.85 ps and 2.45 ps, respectively), while the relative phases remain essentially constant over the displayed distance, thus keeping the in- or out-of-phase character of the bound states, and no energy transfer between the interacting PQSs occurs. We have thus found that the out-of-phase bound state is always the tightest one, with the smallest separation between the interacting solitons. On the other hand, we have found that a similar configuration for the orthogonal case (with the phase difference of $\pi/2$) fail to obtain the same quasi-periodic dynamics as that of the in- and out-of-phase bound states, but can cause their dissolution after collisions. Below, we focus on the out-of-phase bound state cases, in which the interaction produces the strongest effect.

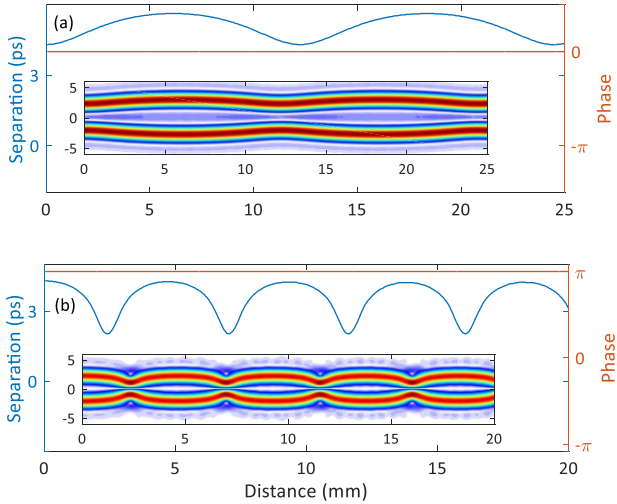

FIG. 5. Typical examples of the oscillatory evolution of the interacting PQSs with the initial separation different from the equilibrium value. (a) The evolutions of their temporal separation and relative phase of the in-phase bound state (keeping the $\Delta\phi=0$ value), as extracted from the numerical simulations. The corresponding profiles of the interacting PQSs are displayed in the inset. (b) The same but for the out-of-phase pair, which keeps the value $\Delta\phi=\pi$ of the relative phase.

Next, we consider out-of-phase pairs of interacting PQSs within a range of different initial separations. Trajectories representing the temporal separation in these oscillatory bound-PQS states were extracted from systematic numerical simulation of Eq. (1). Figure 6 presents a full overview of the evolution of temporal separations in between, exhibiting small-scale oscillations around a certain preferred separation. Here, both the initial velocities and the frequency detuning are set to zero while varying the initial separation, $\Delta\tau_{in}$. In Fig. 6(a), the initial separations exceeding a critical value, 4.85ps, which is indicated by the horizontal red dashed line, give rise to divergent trajectories, isolating the type of bound-PQS pairs (not shown here in detail). Below the critical value, i.e., at $\Delta\tau_{in}$ <4.85ps, all trajectories oscillate around the position specified by the horizontal red solid line, which corresponds to the equilibrium value of the temporal separation for the bound-PQSs state in this case.

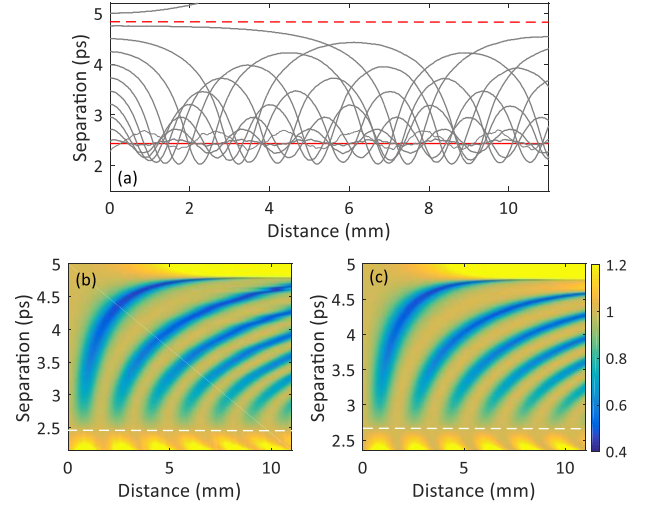

FIG. 6. The separation between the PQSs in the out-of-phase bound states with different values of the initial separation, as a function of the propagation distance, up to 11 mm. (a) Numerical results for the nearly periodic trajectories with the initial separation taking values from 2.25ps to 5ps, with the increment of 0.25ps, illustrating the oscillations around the equilibrium value designated by the horizontal red line. (b, c) Comparison of the interaction in the anti-phase PQS pair, as obtained from numerical simulations (b) and predicted by the interaction model based on Eq. (9) with $\Delta\phi=\pi$ (c), showing good agreement. The color bar represents the normalized value of the separation with respect to the initial one. White horizontal dashed lines in (b, c) designate the equilibrium separation of bound-PQS states and in parallel with the one pointed out by the red solid lines in (a).

To produce the information for their trajectories in a more comprehensive form, the maps of normalized values of the separation, considered as a function of the initial separation $\Delta\tau_{in}$, is plotted in Fig. 6(b), as obtained from a large number of independent simulations. It is seen that the equilibrium separation, designated by the horizontal white dashed line, is close to the position of 2.45ps. The corresponding trajectories for the initial separation $\Delta\tau_{in}$ within the range of [2.45ps, 4.85ps] are quasi-periodic, demonstrating recurrence of bounces, and the periods of bounces increase with the initial separation. The analytical results based on Eq. (9) with $\Delta\phi=\pi$, which are shown in Fig. 6(c), are well matched with their numerical counterparts in Fig. 6(b) in terms of both the equilibrium positions and region of initial separations showing the quasi-periodic behavior. But the quantitative differences exist too, such as a slightly different value of the equilibrium separation, which is close to 2.63ps (showing a difference ≈7.3% from the above-mention numerical value 2.45ps), as indicated by the horizontal white dashed line in Fig. 6(b).

### B. Self-similar structures

The results presented below pertain to the initial configuration in which the constituent pulses may be different from the numerically exact PQSs. In this connection, it is relevant to mention that it was recently reported that PQS pulses with imprecise shapes can also develop into soliton-like modes, whose width and peak power oscillate around certain average value [28]. The respective oscillations of the temporal shape may be identified as a result of the interaction of PQS with the internal mode excited by the imperfect input, cf. Ref. [31]. Thus, this mode offers a new degree of freedom which affects the self-similar interactions between the PQSs, which is outlined as follows below.

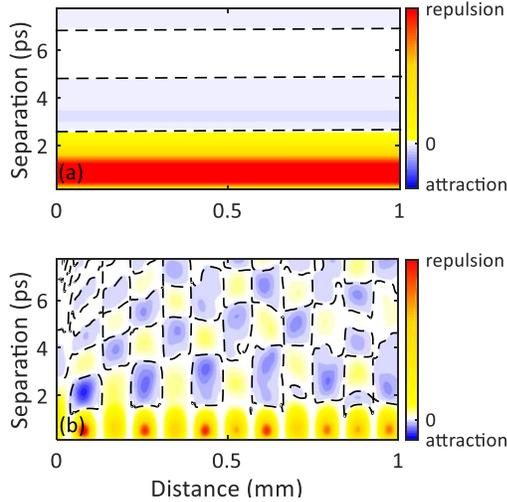

FIG. 7. The dependence of the preceding PQS velocities on their separation within the PQS pair in the course of the propagation over 1mm distance. (a) The ideal PQS bound state. (b) Spatial modulation of the repulsive and attractive regimes for non-ideal PQSs. The color-scales indicate the magnitude of the velocities. The black dashed lines represent the separatrix between the repulsion and attraction.

In the present case, the evolution of bound-PQS states exhibits two spatially periodic (or nearly periodic) phenomena: the slow intrinsic oscillation of the respective PQS pulses and the bouncing recurrence induced by the weak interaction between the pulses. The former effect is naturally associated with the amplitude multiplier $N$ [28], while the latter one is governed by their velocity difference, and then it is determined by the separation between the two PQSs. To illustrate this point, Fig. 7 displays the preceding PQS velocities in the course of the co-propagation of the PQS pair, as a function of the initial separations. Figure 7(a) corresponds to the ideal PQS evolution. On the other hand, for the non-ideal case, doubling the input amplitude ($N=2$) while keeping the width fixed, which mimics the creation of the conventional second-order soliton (breather) [1], drives the PQS pair into a generic oscillatory regime, characterized by a "cranky" form of the respective separatrix net, as shown by black dashed curves in Fig. 7(b). It can be inferred from Fig. 7(b) that the corresponding period of the slow spatial oscillation is ≈0.18mm. In this regime, the velocity difference can be stronger than for the PQSs with the ideal shape.

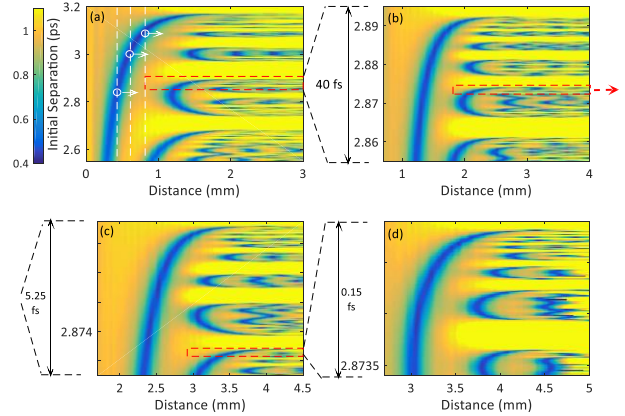

FIG. 8. Self-similar structures of the "tongues" arising from the interaction of well-separated PQSs, showing the fractal nature of the dynamical bound states. (a) The normalized final separation (coded by the color scale) as a function of the propagation distance. This plot is produced by scanning with respect to the initial separation at a fixed total energy. Vertical white dashed lines designate the period of slowly oscillations for a single PQS. White cycles with arrows correspond to the interplay between them and the first bounce, indicating the positions where the "tongues" appear. Panels (b-d) represent a cascaded (four-stage) zoom of the corresponding red boxes, demonstrating the self-similar fractal structure in the parameter plane.

Figure 8(a) shows the results in a format similar to Fig. 6(b), providing a summary of the dynamical behavior of the temporal separation between the interacting PQSs, as produced by the results of the numerical simulations which were collected by varying the initial separation. The color scale is normalized to the initial separation, making it easier to observe the resulting phenomena. In particular, a relatively more complex situation mainly features sustained bounce and fission of the PQS pair at the post-collision stage. This dynamical scenario can be attributed to the interplay of two aspects: nonuniform distribution of relative velocities in the course of the evolution and the nonintegrable of Eq. (1). In this context, the fission of the soliton pair at post-collision stage occurs at the position of the strongest repulsion. As a result, the PQS pair may acquire additional acceleration, and eventually escape from the bounce regime, resulting in permanent dissolution of the bound state. Further, under the condition of the oscillatory propagation of the interacting PQSs, a resonance occurs between the bounce period and the period of the slow oscillations, creating a self-similarity structure. It is clearly observed in Fig. 8(a) that, after the first bounce, several tongues are formed in the parameter plane, alternating between the bound (recurrently bouncing) and unbound (escaping) trajectories, which resembles the general synchronization concept in nonlinear dynamics [38, 39]. The fine structure of these tongues reveals two major points: firstly, the tongue has a distinct shape and appears due to the resonance between the collisional trajectories and slow oscillations, as indicated by a hollow circle with an arrow. In addition to that, multiple sub-tongues are nested inside the main tongue, providing an evident feature of self-similarity.

Figures 8(b)-(c) zoom essential features in the sequential images, such as the existence of the tongues and sub-tongues. The tongues of different orders slightly differ in details, which does not break the reproducibility of the self-similarity structure. The hierarchy of the self-similarity structure is similar to the fractal structure which arises due to the response of nonlinear system to an abrupt change in the parameters in the course of the propagation [40, 41]. The conclusion is that Fig. 8 displays a four stage fractal image. Thus, when the input shapes of PQS pairs deviate from the ideal bound-state form, the parameter range, in which the recurrent bouncing or final fission of the pairs occur, organizes itself into self-similar patterns with fractal characteristics.

## V. CONCLUSIONS

By means of the asymptotic technique, we have derived the system of equations of motion for the temporal positions and relative phase of the pair of weakly interacting PQSs (pure-quartic solitons). The interaction, which may lead to the formation of stationary bound states and oscillating motions of the solitons, is induced by overlapping of their exponentially decaying oscillatory tails. The analysis of the pairwise states reported here can be extended to account for the bound states of three or several solitons.

The internal dynamics of the PQSs pairs includes the formation of stationary bound states and oscillatory ones. The stationary bound states are weakly unstable, hence the oscillatory bound pairs are the most relevant dynamical states in this physical setting. Under certain conditions, the oscillatory bound states eventually split apart. The regions of the periodically oscillating and splitting bound states develop self-similar fractal patterns in the parameter plane (up to the accuracy of the accumulated numerical data). The self-similarity additionally indicates high sensitivity of the evolution on initial condition for the interacting PQS pairs.

The reduction of the full NLSE to the coupled ODE system characterizing the separation and phase dynamics was considered with emphasis on the two-PQS configurations, which, however, can be directly extended to more complex structures, such as systems constituting the sum of more or even infinite well-separated PQSs. Since for each PQS in the multi-pulse configurations we only need to consider its nearest neighboring interactions due to the exponentially decaying tails of PQSs. On the other hand, the approximation of the interaction systems terminating at the term $O(\varepsilon^2)$ makes sense for investigating the weak interaction of bound states within a suitable separation (see comparison of Figs. 6(b) and 6(c)), which, however, may become inaccurate over a wider range of separations, for instance, the extremely weak interaction regimes with larger separations. At this point, it may be necessary to calculate higher-order approximate systems featuring smaller and smaller error terms, forming a hierarchy of structure. In addition, the calculations involving inner products are also subject to machine precision.

## ACKNOWLEDGMENTS

This work was supported by the Guangdong Basic and Applied Basic Research Foundation (2022A1515010326, 2021A1515011532,); Shenzhen Government's Plan of Science and Technology (JCYJ20220818000019040, RCYX20210609103157071); Israel Science Foundation (grant No. 1695/22).

## APPENDIX

### A. The derivation of the system of equations (9) and (10) for the separation and relative phase of the two-soliton state

First, it should be noted that, the weak interaction between the two well-separated PQSs with equal intensities occurs through the overlapping of their tails. When the peak-to-peak separation $\Delta\tau = \tau_2 - \tau_1 > 0$ between them is large, their overlapping quantity $e^{-G\cdot\Delta\tau}$ indicates the weakness of their interaction, namely $e^{-G\cdot\Delta\tau} \sim \varepsilon \ll 1$. Next, using the direct soliton perturbation theory, the soliton parameters, such as the temporal coordinates and the respective phases, slowly evolve over distance $z$. Following the standard multiscale analysis of perturbation, we expand the double-PQS solution of Eq. (1) into the perturbation series:

$$U = e^{i\mu z}(u_1 + u_2 + \varepsilon u^{(1)} + \varepsilon^2 u^{(2)} + \mathcal{O}(\varepsilon^3)) \quad (A1)$$

Here, the function $u_{1,2}$ of the PQS solutions are taken as

$$u_{1,2}(t,z) = u(t-\tau_{1,2})e^{i\phi_{1,2}} \quad (A2)$$

with temporal coordinates $\tau_{1,2}$ and phases $\phi_{1,2}$ being slowly varying functions of distance $z$. Similarly, the corresponding eigenfunctions and adjoint eigenfunctions also have the same temporal and phase shifts. The variables $u^{(1)}, u^{(2)}$ describe the first- and second-order corrections of the linear superposition to the true solution, respectively. We introduce the notations

$$\mathbf{u}^{(n)} = (u^{(n)}(t,z), u^{*(n)}(t,z))^T \quad (A3)$$

Then, by substituting Eq. (A1) in Eq. (1) and separating the terms $\sim \varepsilon$ and $\varepsilon^2$, we obtain the following equations

$$\mathcal{L}(u_1+u_2)\mathbf{u}^{(1)} = -\sum_{j=1}^{2}\mathbf{w}_j\dot{\tau}_j - \sum_{j=1}^{2}\dot{\mathbf{v}}_j\dot{\phi}_j \quad (A4)$$

$$\mathcal{L}(u_1+u_2)\mathbf{u}^{(2)} = -\partial_z\mathbf{u}^{(1)} - \partial_z\mathbf{u}_\Sigma - \mathbf{P} \quad (A5)$$

which $\mathbf{P} = (p(t,z), p^*(t,z))^T = \mathbf{F}(\mathbf{u}_1 + \mathbf{u}_2) - \mathbf{F}(\mathbf{u}_1) - \mathbf{F}(\mathbf{u}_2)$, $p(t,z) = \gamma[(u_1+u_2)|u_1+u_2|^2 - u_1|u_1|^2 - u_2|u_2|^2]$. According to the relationships of eigenfunction, the solution of the linear equation (A4) is

$$\mathbf{u}^{(1)} = -\sum_{j=1}^{2} \mathbf{w}_{jg} \dot{\phi}_j - \sum_{j=1}^{2} \mathbf{v}_{jg} \dot{\tau}_j + \mathcal{O}(\varepsilon^2) \quad (A6)$$

And since the eigenfunctions $\mathbf{w}_j^\dagger, \mathbf{v}_j^\dagger$ correspond to the zero eigenvalue of the adjoint operator $\mathcal{L}^\dagger$, namely $\mathcal{L}^\dagger \mathbf{w}_j^\dagger = \mathcal{L}^\dagger \mathbf{v}_j^\dagger = 0$, so that $\langle \mathcal{L}(u_1+u_2)\mathbf{u}^{(2)}, \mathbf{w}_j^? \rangle = \langle \mathcal{L}(u_1+u_2)\mathbf{u}^{(2)}, \mathbf{v}_j^? \rangle = 0, j=1,2$. By taking the inner product of them with both sides of equation (A5), we can obtain

$$\langle \partial_z \mathbf{u}^{(1)}, \mathbf{w}_j^? \rangle + \langle \partial_z \mathbf{u}_\Sigma, \mathbf{w}_j^\dagger \rangle + \langle \mathbf{P}, \mathbf{w}_j^\dagger \rangle = 0,$$
$$\langle \partial_z \mathbf{u}^{(1)}, \mathbf{v}_j^? \rangle + \langle \partial_z \mathbf{u}_\Sigma, \mathbf{v}_j^\dagger \rangle + \langle \mathbf{P}, \mathbf{v}_j^\dagger \rangle = 0, \quad j=1,2 \quad (A7)$$

By substituting Eq. (A6) in (A7) and using the orthogonal relation $\langle \mathbf{w}_{1,2}, \mathbf{w}_{1,2}^\dagger \rangle = \langle \mathbf{v}_{1,2}, \mathbf{v}_{1,2}^\dagger \rangle = 0$, we can obtain

$$\langle \mathbf{v}_{jg}, \mathbf{v}_j^\dagger \rangle \ddot{\tau}_j = -\langle \mathbf{P}, \mathbf{v}_j^\dagger \rangle, j=1,2 \quad (A8)$$

$$\langle \mathbf{w}_{jg}, \mathbf{w}_j^\dagger \rangle \ddot{\phi}_j - \langle \partial_t \mathbf{v}_{jg}, \mathbf{w}_j^\dagger \rangle (\dot{\tau}_j)^2 = -\langle \mathbf{P}, \mathbf{w}_j^? \rangle, j=1,2 \quad (A9)$$

Obviously, Eqs. (A8) and (A9) are a system of second-order differential equation regarding the soliton parameters such as the time coordinate and phase. The overlap integrals $N_w = \langle \mathbf{w}_g, \mathbf{w}^\dagger \rangle$, $N_v = \langle \mathbf{v}_g, \mathbf{v}^\dagger \rangle$ and $N_{1vw} = \langle \partial_t \mathbf{v}_g, \mathbf{w}^\dagger \rangle$ can be calculated numerically in front of these derivatives. The terms on the right-hand side in Eq. (A8) can be derived as follows:

$$f_{\tau_1} = -\langle \mathbf{v}_1^\dagger \cdot [\mathbf{F}(\mathbf{u}_\Sigma) - \mathbf{F}(\mathbf{u}_1) - \mathbf{F}(\mathbf{u}_2)] \rangle$$
$$= -\langle \mathbf{v}_1^\dagger \cdot [\mathbf{F}(\mathbf{u}_\Sigma) - \mathbf{F}(\mathbf{u}_1)] \rangle - \langle \mathbf{v}_1^\dagger \cdot \mathbf{F}(\mathbf{u}_2) \rangle. \quad (A10)$$

Assuming that the $\tau = 0$ corresponds to the center position between two solitons, i.e., $(\tau_1 + \tau_2)/2 = 0$, the integral interval is separated in two parts:

$$f_{\tau_1} = -\langle \mathbf{v}_1^\dagger \cdot [\mathbf{F}(\mathbf{u}_\Sigma) - \mathbf{F}(\mathbf{u}_1)] \rangle_1 - \langle \mathbf{v}_1^\dagger \cdot \mathbf{F}(\mathbf{u}_2) \rangle_1$$
$$- \langle \mathbf{v}_1^\dagger \cdot [\mathbf{F}(\mathbf{u}_\Sigma) - \mathbf{F}(\mathbf{u}_2)] \rangle_2 - \langle \mathbf{v}_1^\dagger \cdot \mathbf{F}(\mathbf{u}_1) \rangle_2 \quad (A11)$$

where $\langle \cdot \rangle_1 = \int_{-\infty}^{0} (\cdot) dt$ and $\langle \cdot \rangle_2 = \int_{0}^{+\infty} (\cdot) dt$. In the first half-interval $(-\infty, 0]$, where $\mathbf{u}_2$ is small, we obtain

$$\mathbf{F}(\mathbf{u}_\Sigma) - \mathbf{F}(\mathbf{u}_1) \approx \frac{\partial \mathbf{F}}{\partial \mathbf{u}}\Big|_{\mathbf{u}=\mathbf{u}_1} \cdot \mathbf{u}_2, \quad \mathbf{F}(\mathbf{u}_2) \approx \frac{\partial \mathbf{F}}{\partial \mathbf{u}}\Big|_{\mathbf{u}=\mathbf{0}} \cdot \mathbf{u}_2, \quad (A12)$$

In the other half-interval $[0, +\infty)$, where $\mathbf{u}_1$ is small, we obtain

$$\mathbf{F}(\mathbf{u}_\Sigma) - \mathbf{F}(\mathbf{u}_2) \approx \frac{\partial \mathbf{F}}{\partial \mathbf{u}}\Big|_{\mathbf{u}=\mathbf{u}_2} \cdot \mathbf{u}_1, \quad \mathbf{F}(\mathbf{u}_1) \approx \frac{\partial \mathbf{F}}{\partial \mathbf{u}}\Big|_{\mathbf{u}=\mathbf{0}} \cdot \mathbf{u}_1, \quad (A13)$$

Hence, we obtain

$$f_{\tau_1} = -\langle \mathbf{v}_1^\dagger \cdot [\frac{\partial \mathbf{F}}{\partial \mathbf{u}}\Big|_{\mathbf{u}=\mathbf{u}_1} - \frac{\partial \mathbf{F}}{\partial \mathbf{u}}\Big|_{\mathbf{u}=\mathbf{0}}] \mathbf{u}_2 \rangle_1$$
$$- \langle \mathbf{v}_1^\dagger \cdot [\frac{\partial \mathbf{F}}{\partial \mathbf{u}}\Big|_{\mathbf{u}=\mathbf{u}_2} - \frac{\partial \mathbf{F}}{\partial \mathbf{u}}\Big|_{\mathbf{u}=\mathbf{0}}] \mathbf{u}_1 \rangle_2, \quad (A14)$$

In the second half-interval $[0, +\infty)$, where $\mathbf{u}_1$ and $\mathbf{v}_1^\dagger$ are small, the last term on the right-hand side of Eq. (A14) may be neglected. Moreover, considering the adjoint neutral mode $\frac{\partial \mathbf{F}}{\partial \mathbf{u}}\Big|_{\mathbf{u}=\mathbf{u}_1} \mathbf{v}_1^\dagger = 0$, Eq. (A14) is reduced to

$$f_{\tau_1} = \langle \mathbf{v}_1^\dagger \cdot \frac{\partial \mathbf{F}}{\partial \mathbf{u}}\Big|_{\mathbf{u}=\mathbf{0}} \mathbf{u}_2 \rangle_1 \quad (A15)$$

Similarly, the evolution of the temporal positions $\tau_2$ in Eq. (A3) is governed by the equation

$$f_{\tau_2} = \langle \mathbf{v}_2^\dagger \cdot \frac{\partial \mathbf{F}}{\partial \mathbf{u}}\Big|_{\mathbf{u}=\mathbf{0}} \mathbf{u}_1 \rangle_2 \quad (A16)$$

Thus, the right-hand sides in Eqs. (A11) and (A12) can be calculated, and carrying out integration by parts, we obtain:

$$f_{\tau_1, \tau_2} \equiv \langle \mathbf{v}_{1,2}^\dagger \cdot \frac{\partial \mathbf{F}}{\partial \mathbf{u}}\Big|_{\mathbf{u}=\mathbf{0}} \mathbf{u}_{2,1} \rangle_{1,2}$$
$$= 2 \cdot \mathrm{Re} \left\{ \begin{array}{l} i\frac{|\beta_4|}{24}(e^{i(\phi_{2,1}-\phi_{1,2})} + e^{i(\phi_{2,1}-\phi_{1,2})}) \\ \times [\langle v_{1,2}^* \cdot \partial_t^4 u_{2,1} \rangle_{1,2} - \langle u_{1,2}^* \cdot \partial_t^4 v_{2,1} \rangle_{1,2}] \end{array} \right\}$$
$$= \pm 2 \cdot \mathrm{Re} \left\{ \begin{array}{l} i\frac{|\beta_4|}{24}(e^{i(\phi_{2,1}-\phi_{1,2})} + e^{i(\phi_{2,1}-\phi_{1,2})}) \\ \times \left[ \begin{array}{l} v_{1,2}^* \cdot \partial_t^3 u_{2,1} - u_{1,2}^* \cdot \partial_t^3 v_{2,1} \\ - \partial_t v_{1,2}^* \cdot \partial_t^2 u_{2,1} + \partial_t u_{1,2}^* \cdot \partial_t^2 v_{2,1} \end{array} \right]_{t=0} \end{array} \right\}$$

(A17)

Next, applying the symmetry of the PQSs and their neutral modes, namely, $u_0(-t) = u_0(t), v_0(t) = -v_0(-t)$, we obtain

$$f_{\tau_1, \tau_2} = \pm 2 \cdot \mathrm{Re} \left\{ \begin{array}{l} i\frac{|\beta_4|}{24}(e^{i(\phi_{2,1}-\phi_{1,2})} + e^{i(\phi_{2,1}-\phi_{1,2})}) \\ \times \left[ \begin{array}{l} -v_0^* \cdot \partial_t^3 u_0 - u_0 \cdot \partial_t^3 v_0^* \\ -\partial_t v_0^* \cdot \partial_t^2 u_0 - \partial_t u_0 \cdot \partial_t^2 v_0^* \end{array} \right]_{t=\frac{\tau_2-\tau_1}{2}} \end{array} \right\}$$

(A18)

Inserting the asymptotic relations of the trailing tails for the adjoint neutral mode and the PQS solution, we get

$$f_{\tau_1, \tau_2} = \mp 2 \cdot \mathrm{Re} \left\{ \begin{array}{l} i\frac{|\beta_4|}{24} e^{-G(\tau_2-\tau_1)}(e^{i(\phi_2-\phi_1)} + e^{i(\phi_1-\phi_2)}) \\ \times ab^* G^3 (1-i)^3 e^{-i\Gamma(\tau_2-\tau_1)} \end{array} \right\}$$
$$= \mp M_0 e^{-G(\tau_2-\tau_1)} \cos(\phi_2 - \phi_1) \sin[G(\tau_2 - \tau_1) + \Theta]$$

(A19)

where $M_0 = \frac{\sqrt{2}}{6}|\beta_4| \cdot |ab|G^3$, and $\Theta = \arg(ab^*) - 5\pi/4$.

Assuming that the time separation between the interacting PQSs is $\Delta\tau = \tau_2 - \tau_1$ and their phases difference is $\Delta\phi = \phi_2 - \phi_1$, the slow evolution of the time separation obeys the equation

$$\Delta\ddot{\tau} \approx 2M_0 e^{-G \cdot \Delta\tau} \cos(\Delta\phi) \sin(G \cdot \Delta\tau + \Theta). \quad (A20)$$

Considering the differences in the adjoint neutral modes associated with the translational and phase-shifted symmetry, in a similar way, the slow evolution of their phase difference is governed by the following equation:

$$\Delta\ddot{\phi} - Q \cdot \dot{\tau}_{21} \approx -2N_0 e^{-G \cdot \Delta\tau} \sin(\Delta\phi) \cos(G \cdot \Delta\tau + \Xi). \quad (A21)$$

where $\dot{\tau}_{21} = (\dot{\tau}_2)^2 - (\dot{\tau}_1)^2$, $Q = N_{1vw}/N_v$, $N_0 = \frac{1}{N_w}\frac{\sqrt{2}|\beta_4|}{6}|a \cdot c|G^3$, and $\Xi = \arg(ac^*) - \frac{5\pi}{4}$.

B. **The analysis of the weak interaction of PQSs by means of the Manton's method**

Weak interactions between two well-separated PQSs can be also addressed with the help of the method developed by N.

Manton in the terms of a particle model [42]. Recently, Decker et al. applied this method to solve the kink-antikink interactions and its associated dynamics in the $\phi^4$ [43, 44] and NLSE [45] models. Following the pattern of the previous works, here we derive an explicit formula for calculating weak interaction forces induced by the tail-tail overlapping of the PQSs, and recover the equation of motion for the separation of the interacting PQSs.

Firstly, given the translational invariance of PQSs, the conserved total momentum of the field $u$ is introduced:

$$P = \int_{-\infty}^{\infty} \frac{i}{2}(u_t^* u - u^* u_t) dt \quad (B1)$$

To obtain the interaction force between the well-separated PQSs, we differentiate the momentum, within a finite interval $[t_1, t_2]$, with respect to $z$:

$$F = \frac{dP}{dz} = \int_{t_1}^{t_2} \frac{i}{2}(u_{tz}^* u + u_t^* u_z - u_z^* u_t - u^* u_{tz}) dt$$
$$= \int_{t_1}^{t_2} \frac{i}{2}(u_t^* u_z - u_z^* u_t) dt + \frac{i}{2}[u_z^* u - u^* u_z]_{t_1}^{t_2} \quad (B2)$$
$$= [-\frac{|\beta_4|}{24}(u_t^* u_{ttt} - u_{tt}^* u_{tt} + u_{ttt}^* u_t) - \frac{\gamma}{2}|u|^4 + \frac{i}{2}(u_z^* u - u^* u_z)]_{t_1}^{t_2}$$

Based on the expression $u(t,z) = e^{i\mu z} A(t)$, where the real field $A(t)$ is stationary, Eq. (B2) is simplified to

$$F = [\frac{|\beta_4|}{24}(A''^2 - 2A'A''') - \frac{\gamma}{2}A^4 + \mu A^2]_{t_1}^{t_2} = F_{t_2} - F_{t_1} \quad (B3)$$

We consider a symmetrical bimodal structure $A = \eta(t+s) + e^{i\varphi}\eta(t-s)$, consisting of two PQSs $\eta(t)$ centered at $\tau_1 = -s$ and $\tau_2 = s$. By taking $t_1 = 0$ and $t_2 \to \infty$, Eq. (B3) produces the force that the leading soliton exerts on the trailing one. With regard to the fact that $A \to 0$ at $t_2 \to \infty$, the contribution from $t_2$ vanishes. And near $t_1 = 0$, $A$ is the superposition of the tails, giving rise to

$$F \approx [-\frac{|\beta_4|}{24}(A''^2 - 2A'A''') - \mu A^2]_{t=0} \quad (B4)$$

Obviously, the contributions from the self-interaction vanish, hence only the cross-interaction terms produce a non-zero force. Thus, Eq. (B4) can be transformed into

$$F \approx e^{i\varphi}[-\frac{|\beta_4|}{24}(2\eta_1''\eta_2'' - 2(\eta_1'\eta_2''' + \eta_2'\eta_1''')) - 2\mu\eta_1\eta_2]_{t=0} \quad (B5)$$

From Eq. (4), the trailing tail of the leading soliton takes the form of $\eta_1(t+s) = K \cdot (\frac{\mu}{\gamma})^{1/2} e^{-G(t+s)} \cos[G(t+s)+\theta]$, while the leading tail of the trailing soliton is $\eta_2(t-s) = K \cdot (\frac{\mu}{\gamma})^{1/2} e^{G(t-s)} \cos[G(t-s)-\theta]$. Substituting these expressions in Eq. (B5), we obtain

$$F \approx -4e^{i\varphi} K^2 \cdot \mu(\frac{\mu}{\gamma}) e^{-2Gs} \cos(2Gs + 2\theta) \quad (B6)$$

In the case of the weak interaction, the profile of a single PQS is approximated by a static PQS, with the inertial mass given as $M = \int |u|^2 dt \approx \mu$ since the mass is almost conserved during their evolutions. Therefore, the separation $\tau = 2s$ between the interacting solitons obeys the following equation:

$$\ddot{\tau} = -4e^{i\varphi} K^2 \cdot (\frac{\mu}{\gamma}) e^{-G\cdot\tau} \cos(G\cdot\tau + 2\theta) \quad (B7)$$

The equilibrium solution, as well as its local stability, can be predicted by means of Eq. (B7). For example, the equilibrium $\tau = 2.6313 (4.753798)$ ps is stable for the out-of-phase (in-phase) case. Further, to solve Eq. (B7), we employed the MATLAB's built-in ODE solver, ode15s, with the initial condition of $\tau_0 = 4.4739$ ps and $\dot{\tau}_0 = 0$ (initial zero velocity), which correspond to the two cases shown above in Fig. 5. The so-obtained phase portrait is displayed in Fig. 9(a). The closed orbit, represented by the blue curve, surrounds the equilibrium $\tau = 4.753798$ ps, which corresponds to the in-phase case, while the red orbit surrounds the equilibrium $\tau = 2.6313$ ps, which corresponds to the out-of-phase case. For comparison, the corresponding results extracted from PDE simulations, which are displayed above in Fig. 5, are shown in Fig. 9(b). Good consistency is observed, while a difference is seen too. This can be explained as follows: the ODE, compatible with the energy conservation, can only lead to the mutual rebound of the interacting PQSs, while the PDE simulations of the present non-integrable system may transfer the energy from the PQS translational motion to the internal or radiation modes. This dynamical scenario can excite the radiation of energy, and thus affect the soliton's evolution. In addition, higher-order contributions to the tail-mediated interaction are neglected in the derivation of ODE (B7), which can also lead to some discrepancy with the PDE simulations.

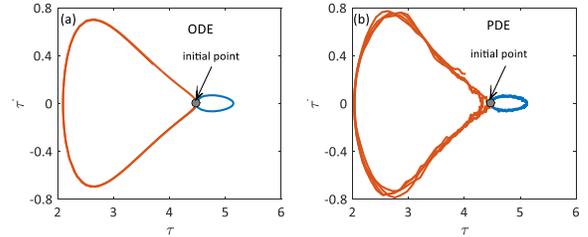

FIG. 9. The comparison between the results (left column) produced by ODE (B7) and the corresponding results (right column) produced by the full PDE simulation in Fig. 5. The blue and red curves correspond to the in-phase and out-of-phase case, respectively.

Both the effective particle model (Eq. (B7)) and the direct perturbation analysis presented in Appendix A are used to address the interaction of the weak-overlapping PQSs. The former method uses the dynamical invariants (mass and momentum) to derive the interaction forces and evolution equation for the separation between the two PQSs. This method is simple but has some limitations, such as the inability to take into account the emission of radiation, which is the by-product of soliton collisions in the non-integrable model. On the other hand, the radiation can be taken into regard by the linearized solution through integrals of its complete eigenfunctions, as shown in Appendix A.